\documentclass[aip, amsmath, amssymb, preprint]{revtex4-1}

\usepackage[T1]{fontenc} 

\usepackage{amsmath,color}
\usepackage{amssymb}
\usepackage{amsfonts}
\usepackage{amsthm}
\usepackage{mathtools}

\usepackage[title]{appendix}

\usepackage{algorithm}
\usepackage{algpseudocode}
\usepackage{dsfont}

\usepackage{bbold}
\usepackage{chemformula}

\DeclarePairedDelimiter\bra{\langle}{\rvert}
\DeclarePairedDelimiter\ket{\lvert}{\rangle}
\DeclarePairedDelimiterX\braket[2]{\langle}{\rangle}{#1 \delimsize\vert #2}
\DeclarePairedDelimiterX\braket3[3]{\langle}{\rangle}{#1 \delimsize\vert #2 \delimsize\vert #3}

\newcommand{\vP}{{\boldsymbol{\mathcal{P}}}}

\newcommand{\vE}{\mathbf{E}}

\newcommand{\vB}{\mathbf{B}}

\newcommand{\vD}{\mathbf{D}}

\newcommand{\vmu}{\boldsymbol{\mu}}

\newcommand{\vR}{\mathbf{r}}

\newcommand{\vk}{\mathbf{k}}

\newcommand{\vxi}{\boldsymbol{\xi}}

\newcommand{\hH}{\hat{H}}

\newcommand{\hE}{\hat{\mathbf{E}}}

\newcommand{\hA}{\hat{\mathbf{A}}}

\newcommand{\tr}[1]{\text{Tr}\left(#1\right)}
\newcommand{\trtwo}[2]{\text{Tr}_{\text{#1}}\left(#2\right)}

\newcommand{\avg}[1]{\left\langle #1\right\rangle}
\newcommand{\red}[1]{{\color{black} #1}}

\begin{document}

	\title{On the Origin of Ground-State Vacuum-Field Catalysis: Equilibrium Consideration}%
	
	\author{Tao E. Li}%
	\email{taoli@sas.upenn.edu}
	\affiliation{Department of Chemistry, University of Pennsylvania, Philadelphia, Pennsylvania 19104, USA}

	
	\author{Abraham Nitzan} 
	\email{anitzan@sas.upenn.edu}
	\affiliation{Department of Chemistry, University of Pennsylvania, Philadelphia, Pennsylvania 19104, USA}
	\affiliation{School of Chemistry, Tel Aviv University, Tel Aviv 69978, Israel}
	
	\author{Joseph E. Subotnik}
	\email{subotnik@sas.upenn.edu}
	\affiliation{Department of Chemistry, University of Pennsylvania, Philadelphia, Pennsylvania 19104, USA}

	\begin{abstract}
		Recent experiments suggest that vibrational strong coupling (VSC) may significantly modify ground-state chemical reactions and their rates even without external pumping. The intrinsic mechanism of this "vacuum-field catalysis" remains largely unclear. Generally, modifications of thermal reactions in the ground electronic states \red{can be} caused by equilibrium or non-equilibrium effects. The former are associated with modifications of the reactant equilibrium distribution as expressed by  the transition state theory of chemical reaction rates, while the latter stem from the dynamics of reaching and leaving transition configurations. Here, we examine the VSC effect  in a cavity environment on chemical rates as calculated by transition state theory. Our approach is to  examine the effect of coupling to cavity mode(s) on the potential of mean force (PMF) associated with the reaction coordinate. \red{Within the context of classical nuclei and classical photons,}  we find that while the PMF can be affected by the cavity environment, this effect is negligible for the usual cavities  used to examine VSC situations.
	\end{abstract}

	\maketitle
	
	
	\section{Introduction}
	\label{sec:intro}
	Strong light-matter interactions can significantly modify the intrinsic properties of matter by forming hybrid light-matter states (i.e., polaritons)\cite{Herrera2019}. 
	The most common manifestation of this interaction is the Rabi splitting\cite{Kaluzny1983,Raimond2001,Torma2015,Chikkaraddy2016},
	\begin{equation}\label{eq:RabiSpliting}
	\Omega_N = 2\sqrt{N}g_0 ,
	\end{equation}
	where $N$ is the number of closely spaced (relative to the wavelength) molecules, resulting from the coupling $g_0$  between the molecular  transition and a resonant cavity mode.
	Moreover, due to the modification of the molecule's electronic levels,  properties such as the energy transfer rate\cite{Zhong2016}, conductivity\cite{Orgiu2015}, and the photochemical (i.e., light induced chemical change) reaction rates\cite{Hutchison2012,Herrera2016,Galego2016,Mandal2019} can also be changed by strong light-matter interactions. 
	The Rabi splitting in Eq. \eqref{eq:RabiSpliting} was recently observed\cite{Long2015,Chervy2018} in the case of vibrational strong coupling (VSC) --- when an infrared cavity mode is resonantly coupled to a molecular vibrational mode. More intriguingly, it has been argued that the effect of VSC on the ground-state potential energy surface (PES) can result in an N-dependent modification of chemical reaction rates in the absence of external pumping, an effect termed vacuum-field catalysis\cite{Hiura2018,Hiura2019}. Indeed, such behaviors were  reported recently\cite{Thomas2016,Lather2019,Hiura2018,Thomas2019_science,Vergauwe2019}.
	
	This effect has attracted great attention of late as a novel manifestation of strong molecules--radiation field coupling with possible implications for catalysis.  However, the physics underlying this behavior remains unclear. The difficulty is illustrated\cite{Imran2019,Xiang2018,Pino2015,Campos-Gonzalez-Angulo2019} by considering the following quantum Hamiltonian\cite{Hernandez2019,FriskKockum2019,Du2018} used to analyze VSC effects (we set $\hbar = 1$):
	\begin{equation}\label{eq:H_quantum_rwa}
		\hat{H}_{\text{RWA}} = \omega_0 \hat{a}^{\dagger}\hat{a} + \sum_{i=1}^{N}\omega_0 \hat{b}_i^{\dagger}\hat{b}_i + \sum_{i=1}^{N}g_0(\hat{a}\hat{b}_i^{\dagger} + \hat{a}^{\dagger}\hat{b}_i)
	\end{equation}
	where $\omega_0$ denotes the energy of the cavity mode and molecular vibrations (assuming all are at resonance), and $\hat{a}$ ($\hat{a}^{\dagger}$) and $\hat{b}_i$ ($\hat{b}_i^{\dagger}$) denote the creation (annihilation) operators for the cavity mode and the $i$-th vibrational mode, respectively. Note that the rotating-wave approximation  (RWA) is taken in Eq. \eqref{eq:H_quantum_rwa}, which is a good approximation at resonance and when $\Omega_{N} \ll \omega_{0}$. Diagonalizing Eq. \eqref{eq:H_quantum_rwa} leads to a pair of polariton states  [known as the upper (UP) and lower (LP) polaritons]  with  frequency difference $\Omega_{N}$ between them,
	and the remaining $N-1$ quantum states (known as the dark states) are totally decoupled from the cavity mode. 
	Excitation from the ground state populates the polaritonic modes, however it is not clear that these modes can affect the rate of a process in which a single molecule undergoes a chemical change. 
	Put differently, at thermal equilibrium most molecules are in the ground state and most of those with thermal energy of the order  $\hbar\omega_{0}$  populate dark states, evolving just like bare molecules.
	Such a picture would not seem to agree with the experiments described above and although a few theoretical studies have investigated VSC assisted vacuum-field catalysis\cite{Galego2019,Campos-Gonzalez-Angulo2019,Hiura2019}, to the best of our knowledge, a convincing and universal theory for the dependence on molecular density is still unavailable.
	 For example, if one posits that an unknown mechanism were to force  the UP or LP states to be a doorway to a chemical reaction, then the activation energy change should shift linearly with $\Omega_{N}$\cite{Hiura2019}, in contrast with recent studies demonstrating that both the entropy and enthalpy of chemical reactions vary \textit{nonlinearly} as a function of $\Omega_{N}$\cite{Thomas2019}.
	 
	 Another way to look at the cavity effect on chemical reactions is to adhere to standard chemical rate theory when considering the possible effect of multimolecular coupling to cavity mode(s) on the chemical rate. Recent experiments indicate that "vacuum-field catalysis" is a collective effect, sometimes involving a macroscopic number of molecules\cite{Thomas2016,Lather2019,Hiura2018,Thomas2019_science}. For such system sizes any full quantum treatment beyond the harmonic level is computationally very expensive. However, as often observed in chemical rate calculations, a classical picture should already contain the essence of the effect, with quantum effects providing additional corrections.

	Chemical rate processes can be observed in different regimes. The simplest and most widely used picture is transition state theory (TST), which rests on the assumption that reactant(s) maintain an equilibrium distribution in their configuration/velocity space even under conditions of chemical non-equilibrium. In its classical unimolecular form, this theory assumes that a molecule undergoes the considered chemical change when its reaction coordinate $x$ reaches a certain position ("transition configuration"). The probability to reach this configuration is determined, according to the equilibrium assumption, by the potential of mean force  (PMF, \cite{Kirkwood1935,Tuckerman2010}) or free energy function, $F(x)$, defined by
	\begin{widetext}
	\begin{equation}\label{eq:PMF}
	\exp\left[-\beta F(x) \right] \equiv \int d\vR^N \int d \mathbf{p}^N \delta\left(x - \bar{x}(\vR^N)\right) \exp\left[-\beta H(\mathbf{p}^N, \vR^N)\right]
	\end{equation}
	\end{widetext}
	where $\beta = (k_B T)^{-1}$, $\vR^N = (\vR_1, \cdots, \vR_N)$ is the configuration vector of the $N$-particle system and  $x = \bar{x}(\vR^N)$ defines the reaction coordinate. At this level of description, assuming that the molecular reaction coordinate itself is the same in and out of the cavity, the cavity effect on the reaction rate amounts to a change in  the PMF and in particular the barrier height (i.e., activation energy) for the relevant potential. 
	
	In this paper we investigate the implications of these considerations, as well as some of their quantum counterparts, for possible cavity effects on chemical rates. \red{Within the context of classical nuclei and classical photons,}  we find that the PMF associated with the reaction coordinate of a single molecule can be modified by the cavity environment. However, \red{the} dependence of this modification on the number of molecules interacting with the cavity can result only from the cavity effect on intermolecular interactions. Such cavity effects are not necessarily related to the resonant interactions with a cavity mode, and are identical to similar effects previously studied using image potential considerations\cite{Takae2013,DeBernardis2018,Kobayashi1995,Hopmeier1999,Goldstein1997}. 
	\red{Admittedly, there is one nuance here: one can separate the effects of intermolecular interactions from the effects of intramolecular interactions rigorously only within classical mechanics. Nevertheless, from the treatment below, we conclude} that VSC catalysis cannot be directly explained through static equilibrium considerations \red{when classical nuclei and photons are assumed}.
	It should be emphasized that this conclusion does not preclude possible cavity effects \red{as originating from} inherently non-equilibrium effects that can dominate chemical rates in other dynamical regimes\red{;  as will be shown below, it is also possible  that quantum mechanical effects could in principle lead to some collective effects on the PMF --- though for now we will hypothesize that such collective effects are unlikely. }
	
	A specific outline of the paper below is as follows.
	In Sec. \ref{sec:PMF} we consider the implication of the interaction of $N$ identical molecules with the cavity field \red{\textit{under the assumption that the cavity environment does not modify the direct intermolecular Coulombic interactions on the PMF associated with any single molecule}}. Working in the Coulomb gauge, we show that while the cavity environment does modify the single-molecule PMF, this modification does not depend on $N$. This observation should not of course be gauge dependent and indeed we validate this observation in the equivalent dipole gauge (Power--Zienau--Woolley Hamiltonian\cite{Babiker1983,Cohen-Tannoudji1997}) representation which provides a somewhat different perspective of this issue. Now while disregarding cavity effects on the intermolecular Coulombic interactions may sometimes be a good approximation, it is fundamentally inconsistent with the theory of light-matter interaction, where a proper balance between such interactions and those mediated by the longitudinal part of the radiation field is required for achieving a fully retarded character of these interactions. Thus, in Sec. \ref{sec:discussion}  we reexamine this issue of intermolecular interactions by using a reasonable set of parameters, and conclude that the cavity effects on intermolecular interactions are too small to explain observations of VSC-induced collective (namely $N$-dependent) effects on chemical rates \red{(at least within the context of classical nuclei and photons)}. \red{Finally, in the Appendix, we address the question of the PMF in the context of quantum nuclei and photons using a path integral approach, but we are unable to establish with complete rigor whether the PMF is modified in any collective way; future numerical work will be needed in this regard \cite{Li2020water}.}

	\section{Potential of Mean Force}
	\label{sec:PMF}
	
	\subsection{Single-cavity-mode approximation}
	We start with the standard Coulomb-gauge expression for the Hamiltonian of a system of charges interacting with the radiation field represented by a single cavity mode of frequency $\omega$ (the formulation remains the same if a sum over cavity modes is taken)
	\begin{equation}\label{eq:HC}
	\hH = \sum_{i} \frac{1}{2m_i}\left[\hat{\mathbf{p}}_i - Z_i e \hA(\hat{\vR}_i)\right]^2 
	+ \hat{V}_{\text{Coul}}\left(\{\hat{\vR}_i\}\right) + \hbar\omega \hat{a}^{\dagger}\hat{a}
	\end{equation}
	Here $\hat{a}^{\dagger}$ ($\hat{a}$) denote the creation (annihilation) operator of the cavity mode,  $e$ denotes the electronic charge, $m_i$, $\hat{\mathbf{p}}_i$, $\hat{\vR}_i$ are the mass, momentum and position of the $i$-particle of charge $Z_i e$, respectively. 
	$\hat{V}_{\text{Coul}}\left(\{\hat{\vR}_i\}\right)$ denotes the electrostatic interaction between all charged particles and
	\begin{equation}\label{eq:A_def}
	\hA(\vR) = \sqrt{\frac{\hbar}{2\omega \Omega \epsilon_0}} \vxi 
					\left[
					\exp(i \vk \cdot \vR) \hat{a} +
					\exp(-i \vk \cdot \vR) \hat{a}^{\dagger}
					\right]
	\end{equation}
	denotes the vector potential of the electromagnetic field. $\epsilon_0$ is the vacuum permittivity  while $\Omega$, $\vxi$, and $\vk$ denote the cavity volume, mode polarization, and wave vectors that satisfy $\vxi \cdot \vk = 0$. Next, we make the long-wave approximation, assuming that the size of the molecular ensemble is much smaller then the wavelength of the cavity mode (or when many modes are considered --- of all relevant modes). In this case Eq. \eqref{eq:A_def} can be approximated by
	\begin{equation}\label{eq:A_long_wave}
	\hA(\mathbf{0}) = \sqrt{\frac{\hbar}{2\omega \Omega \epsilon_0}} \vxi 
	\left( \hat{a} + \hat{a}^{\dagger} \right)
	\end{equation}
	The molecular system is characterized by the total dipole moment
	\begin{equation}
	\hat{\vmu}_S = \sum_{i} Z_i e \hat{\vR}_i
	\end{equation}
	which may be also grouped into the dipole moments of the individual molecules (indexed by $n$)
		\begin{equation}
	\hat{\vmu}_S = \sum_n \hat{\vmu}_n;  \ \ \   \hat{\vmu}_n = \sum_{j \in n} Z_j e \hat{\vR}_j
		\end{equation}
	The molecules are assumed neutral, $\sum \limits_{j \in n} Z_j = 0$, hence $\hat{\vmu}_n$ does not depend on the choice of origin of coordinates. Next we perform the unitary G\"oppert-Mayer transformation, $\hat{H}' = \hat{U} \hat{H} \hat{U}^{\dagger}$ with $\hat{U} = \exp\left[-\frac{i}{\hbar} \hat{\vmu}_S\cdot \hA(\mathbf{0})\right]$ leading to
	\begin{equation}\label{eq:H'}
	\begin{aligned}
		\hat{H}' &= \sum_{i} \frac{\hat{\mathbf{p}}_i^2}{2 m_i} +  \hat{V}_{\text{Coul}}\left(\{\hat{\vR}_i\}\right) \\
		&
	+ \hbar\omega \hat{a}^{\dagger}\hat{a} - \hat{\vmu}_S\cdot \hE + \frac{1}{2\Omega\epsilon_0}\left(\hat{\vmu}_S \cdot \vxi\right)^2
	\end{aligned}
	\end{equation}
	where $\hE = i \vxi \sqrt{\frac{\hbar\omega}{2\Omega\epsilon_0}}\left(\hat{a} - \hat{a}^{\dagger}\right)$ is the operator representing the electric field associated with the cavity mode.
	
	Anticipating the possibility of making a classical approximation, it is useful also to recast the photon operators $\hat{a}$ and $\hat{a}^{\dagger}$ in Eq. \eqref{eq:H'} in terms of coordinate and momentum operators. Putting $\hat{a} = (2\hbar\omega)^{-1/2}(\omega \hat{q} - i \hat{p})$ and $\hat{a}^{\dagger} = (2\hbar\omega)^{-1/2}(\omega \hat{q} + i \hat{p})$ leads to $\hbar\omega \hat{a}^{\dagger}\hat{a} = \frac{1}{2}(\omega^2 \hat{q}^2 + \hat{p}^2)$ and  $\hat{\vmu}_S\cdot \hE = (\Omega\epsilon_0)^{-1/2}\left(\hat{\vmu}_S\cdot \vxi\right) \hat{p}$,  the Hamiltonian  \eqref{eq:H'} becomes
	\begin{equation}\label{eq:H'2}
	\begin{aligned}
		\hat{H}' &=  \sum_{i} \frac{\hat{\mathbf{p}}_i^2}{2 m_i} +  \hat{V}_{\text{Coul}}\left(\{\hat{\vR}_i\}\right) \\
		&
	+ \frac{1}{2}\omega^2 \hat{q}^2 + \frac{1}{2}\left(\hat{p} - \sqrt{\frac{1}{\Omega\epsilon_0}}\left(\hat{\vmu}_S\cdot \vxi \right) \right)^2
	\end{aligned}
	\end{equation}
	More generally, the last two terms of Eq. \eqref{eq:H'2} will be summed over all relevant cavity modes. Our focus is on effects of vibrational strong coupling, where relevant cavity modes are assumed to evolve on timescales similar to molecular vibrational motions. The effect of faster modes, which under the Born--Oppenheimer timescale separation should be considered together with the electronic Hamiltonian, will be disregarded assuming that their energetic consequences (for electromagnetic vacuum) are small relative to electronic energy scales.

	With this assumption we may proceed to consider the potential \red{energy surface associated with} the ground electronic state of the Hamiltonian \eqref{eq:H'2} assuming that the photon dynamics, namely the evolution  $(q(t), p(t))$  takes place on a nuclear timescale. Disregarding intermolecular interactions embedded in  $ \hat{V}_{\text{Coul}}\left(\{\hat{\vR}_i\}\right)$, the ground-state nuclear/cavity photon Hamiltonian is
	\begin{equation}\label{eq:Hcl}
	H_{\text{nuc}} = \sum_{n=1}^{N} \frac{P_n^2}{2M_n} + \frac{1}{2}\omega^2 q^2 + V(\{R\red{_n}\}, p) 
	\end{equation}
	where we now use capital $P\red{_n}$ and $R\red{_n}$ (rather than $\hat{\mathbf{p}}_i$ and $\hat{\mathbf{r}}_i$) to represent nuclear degrees of freedom \red{(DoFs)}.
	 The potential surface (electronic energy) for the \red{$N$-molecule aggregate plus photon is:}
	\begin{equation}\label{eq:PES}
	\begin{aligned}
		V(\{R\red{_n}\}, p) &= \sum_{n = 1}^{N} E_g(R_n) \\
		&+ \frac{1}{2}\avg{\Psi_G \Bigg | \left[ p - \sqrt{\frac{1}{\Omega \epsilon_0}}\left( \sum_{n=1}^N\hat{\vmu}_n\cdot \vxi\right)\right]^2 \Bigg |\Psi_G}
	\end{aligned}
	\end{equation}
	Here, $E_g(R_n)$, the electronic ground state energy of an individual molecule, is a function of its nuclear configuration (represented by $R\red{_n}$, with $\{R\red{_n}\}$ denoting the nuclear configurations of all $N$ molecules), $P_n (n = 1, \cdots, N)$ denotes the nuclear momentum and \red{$\Psi_G\left(\{R\red{_n}\}\right)$ denotes the ground-state electronic wave function for the molecular subsystem}.  Note that the designation of $q$ or $p$ as photon coordinate and momentum or vice-verse is immaterial. Also note that for each molecule the dipole operator $\hat{\vmu}_{n} = \hat{\vmu}_{ne} + \hat{\vmu}_{n, nuc}$  is a sum of electronic and nuclear terms, but these two terms should be considered together, otherwise each will depend on the choice of origin.
	
	The second term in Eq. \eqref{eq:PES} should be handled with care because the square introduces bimolecular terms.  \red{According to the  Hartree approximation, let $\Psi_G\left(\{R\red{_n}\}\right) = \prod_{n=1}^{N}\psi_{ng}(R_n)$ be a product of single-molecule electronic ground states, and also denote $\hat{d}_n \equiv \hat{\vmu}_n\cdot \vxi$ and $d_{ng} \equiv \avg{\psi_{ng}(R_n)|\hat{\vmu}_n |\psi_{ng}(R_n)}\cdot \vxi$. The second term in Eq. \eqref{eq:PES} becomes}
	\begin{widetext}
	\begin{equation}\label{eq:Treatment_2}
	\begin{aligned}
	& \red{\frac{1}{2}}\avg{\Psi_G \Bigg | \left[ p - \sqrt{\frac{1}{\Omega \epsilon_0}}\left( \sum_{n=1}^N\hat{d}_n  \right)\right]^2 \Bigg |\Psi_G}  \\
	&=
	\red{\frac{1}{2}}\left[ p - \sqrt{\frac{1}{\Omega \epsilon_0}} \left( \sum_{n=1}^N d_{ng}(R_n)  \right) \right]^2 
	+ \frac{1}{\red{2}\Omega\epsilon_0} \left[ \avg{\Psi_G \Bigg |\sum_{n = 1}^{N} \left(\hat{d}_n - d_{ng}(R_n)\right)^2 \Bigg | \Psi_G}  \right]
 \\
	&= \red{\frac{1}{2}\left[ p - \sqrt{\frac{1}{\Omega \epsilon_0}} \left( \sum_{n=1}^N d_{ng}(R_n)  \right) \right]^2 
	+ \sum_{n = 1}^{N}  \frac{1}{2\Omega\epsilon_0}  \left[ \avg{\psi_{ng} \Bigg |\left(\hat{d}_n - d_{ng}(R_n)\right)^2 \Bigg | \psi_{ng}}  \right]}
		\end{aligned}
	\end{equation}
	\end{widetext}
	\red{Note that in calculating Eq. \eqref{eq:Treatment_2}, the main issue is how to calculate the expectation value of the self-dipole term $\braket3{\Psi_G}{\left(\sum_n \hat{d}_n\right)^2}{\Psi_G} = \sum_{n, l}\braket3{\Psi_G}{\hat{d}_n\hat{d}_l}{\Psi_G}$. For $n=l$, the expression becomes $\braket3{\psi_{ng}}{\hat{d}_n^2}{\psi_{ng}} = d_{ng}^2 + \braket3{\psi_{ng}}{(\hat{d}_n -  d_{ng})^2}{\psi_{ng}}$. For $n\neq l$,  the expression becomes $\braket3{\psi_{ng}\psi_{lg}}{\hat{d}_n \hat{d}_l}{\psi_{ng}\psi_{lg}}  = d_{ng}d_{lg}$.
	
}

	\red{We then substitute Eq. \eqref{eq:Treatment_2}} in Eqs. \eqref{eq:PES} and \eqref{eq:Hcl} to calculate the the \red{classical} PMF according to (following Eq. \eqref{eq:PMF})
	\begin{equation}\label{eq:PMF2}
	e^{-\beta F(R_j)} \equiv \int d \{P\}  d '\{R\} dp dq  e^{-\beta H_{\text{nuc}}(\{P\red{_n}\}, \{R\red{_n}\}, p, q) }
	\end{equation}
	where $d '\{R\red{_n}\}$ denotes integration the nuclear coordinates of all molecules except molecule $j$.  \red{Note that in evaluating Eq. \eqref{eq:PMF2},}
	the contribution from the first term on the bottom line of  Eq. \eqref{eq:Treatment_2} will 
	effectively disappear as it will be incorporated within  the $p$ integral\red{, i.e., $\int_{-\infty}^{+\infty} dp \ \exp\left(-\beta\left[ p - \sqrt{\frac{1}{\Omega \epsilon_0}} \left( \sum_{n=1}^N d_{ng}(R_n)  \right) \right]^2/2\right) = \int_{-\infty}^{+\infty} dp \ \exp\left(-\beta p^2/2\right) = \text{const}$}.  After the integrations over $p$, $q$, and $\{P\red{_n}\}$, Eq. \eqref{eq:PMF2} becomes (up to a constant)
\begin{widetext}
	\begin{equation}\label{eq:PMF_15}
	\begin{aligned}
		e^{-\beta F(R_j) } & \sim \int d '\{R\red{_n}\} \prod_{n=1}^{N} \exp\left[-\beta \left(
	E_g(R_n) + \frac{1}{2\Omega\epsilon_0}\avg{\psi_{ng}(R_n)\Bigg |
		\left(\hat{d}_n - d_{ng}(R_n)\right)^2 
		\Bigg | \psi_{ng}(R_n)} \right)\right]  \\
	&\sim 
	\exp\left[-\beta \left(
	E_g(R_j) + \frac{1}{2\Omega\epsilon_0}\avg{\psi_{jg}(R_j)\Bigg |
		\left(\hat{d}_j - d_{jg}(R_j)\right)^2 
		\Bigg | \psi_{jg}(R_j)} \right)\right] 
	\end{aligned}
	\end{equation}
	\end{widetext}
	We find an effect on the PMF for a single molecule is
	\begin{equation}
	E_g(R) \rightarrow E_g(R) + \frac{1}{2\Omega\epsilon_0} \delta d^2(R)
	\end{equation}
	\red{Here, the self-dipole fluctuation term is} $\delta d^2(R) \equiv \avg{\psi_{g}(R)\Big |
		\left(\hat{d} - d_{g}(R)\right)^2 
		\Big | \psi_{g}(R)}$ and $d_{g}(R) = \avg{\psi_{g}(R) \Big |\hat{d} \Big |\psi_{g}(R)}$. 
	
	\red{The following points should be noted. First, for sufficiently small cavities ($\Omega \rightarrow 0$), the term $\frac{1}{2\Omega\epsilon_0} \delta d^2(R)$ can significantly modify the PMF and should be able to modify the ground-state chemistry. However, such an effect is not "collective" (that is $N$-dependent), and for typical VSC experiments where micron-length cavities are used, the term $\frac{1}{2\Omega\epsilon_0} \delta d^2(R)$ should be negligible.}

	\red{Second, a word is now appropriate vis-a-vis quantum versus classical mechanics. The term $\frac{1}{2\Omega\epsilon_0} \delta d^2(R)$ has a purely quantum origin and arises from the quantum treatment of the electron. When the molecules are treated entirely classically, the fluctuation $\delta d^2(R)$ would be zero and so the PMF would remain completely unchanged by the cavity. 
	Such behavior is often discussed in the context of magnetism (denoted by $\mathbf{m}$) where the role of classical vs quantum mechanics is paramount \cite{Nolting2009}. According to the Boh--van Leeuwen theorem, when molecules are treated classically, the evaluation of the momentum integral for the thermal-averaged magnetism (in a similar way as calculating Eq. \eqref{eq:PMF2}) leads to the conclusion $\avg{\mathbf{m}} =0$ and therefore one would predict a lack of magnetism. Of course, by symmetry, a fully quantum treatment of magnetism (that includes exchange) must also yield $\avg{\mathbf{m}} =0$, but in practice this result  represents the average of two stable, symmetry-broken solutions with plus and minus magnetic moments. Thus, given the important distinction between classical and quantum mechanics, one may question our finding for the PMF above: even though we allowed for quantum \textit{electrons}, we did rely on a classical assumption of \textit{nuclear} and \textit{photonic} DoFs. To that end, in the Appendix, we will present  initial results for the fully quantum case as well.}

	\red{Third and finally, yet another word of caution is also appropriate at this time.  While the Hamiltonian \eqref{eq:Hcl} and \eqref{eq:PES} is often used to discuss molecular cavity QED phenomena, it constitutes an approximation whose consistency has yet to be checked. In principle, within an  exact quantum Hamiltonian for a system with light-matter interactions, all instantaneous Coulombic intermolecular interactions are canceled exactly by the presence of terms that arise from the self-interaction of dipoles arising from the last term in Eq. \eqref{eq:PES}. This exact cancellation allows for causality to be enforced such that all intermolecular interactions are carried by the transverse photon field at the speed of light. However, to achieve this cancellation one should sum contributions from all cavity modes. The approximation embedded in the Hamiltonian \eqref{eq:HC}-\eqref{eq:H'2}, like in many other treatments of light-matter interactions in optical cavities, is to consider only such cavity mode(s) that is (are) close to resonance with the molecular vibration(s) while keeping the intermolecular electrostatic interactions intact. While such an approximation appears reasonable when retardation effects are unimportant, it leaves open consistency issues that might arise in such systems.  We note in passing that these intermolecular interactions  (which are included within $\hat{V}_{\text{Coul}}$ in Eq. \eqref{eq:H'2})   are absent from Eq. \eqref{eq:Hcl}, and in principle should be included, but we argue below that their dependence on the cavity environment is expressed only in cavity sizes much smaller than those actually used in the experimental work on VSC effects on ground-state chemical processes.
	Vice versa, if we did include these interactions, they will appear in the classical PMF in Eq. \eqref{eq:PMF_15}, but their form would not be altered by integration over the photon coordinates. For all of these reasons, we disregard them here.	}
	
	\subsection{Full consideration of causality}
	
	A somewhat different perspective on \red{the third point above (i.e. the question of a proper light-matter Hamiltonian)} can be obtained by examining the problem using another popular gauge, the dipole (or Power--Zienau--Woolley) gauge. This representation is particularly convenient when the system under study comprises neutral units (molecules), well separated relative to their size, that are characterized by their charge distributions, in particular their dipoles. 
	
	Accordingly (and unlike Eq. \eqref{eq:HC} that starts from individual charged electrons and nuclei) our starting point focuses on such a system, and as before we restrict ourselves to a classical description corresponding to the high temperature limit for the time and energy scales associated with nuclear  motions and the corresponding electromagnetic modes. The Hamiltonian is  taken to be\cite{Cohen-Tannoudji1997}
	\begin{equation}\label{eq:H_PZW}
	\begin{aligned}
		H &= \sum_{n = 1}^{N} H_n + \sum_{n < l} V_{\text{Coul}}^{(nl)} \\
		&+ \frac{1}{2} \int d \vR \left(\frac{1}{\epsilon_0} \vD(\vR)\cdot \vD(\vR) + \frac{1}{\mu_0} \vB(\vR)\cdot \vB(\vR)\right) \\
		&
	- \frac{1}{\epsilon_0}\int d\vR \vD(\vR)\cdot \vP_{\perp}(\vR) + \frac{1}{2\epsilon_0} \int d\vR \vP_{\perp}(\vR)\cdot \vP_{\perp}(\vR)
	\end{aligned}
	\end{equation}
	where $H_n$ is the Hamiltonian for the $n$-th molecule and $V_{\text{Coul}}^{(nl)}$ are the Coulombic interactions between molecules that we assume to be dominated by dipole-dipole interactions\cite{footnote1}, $\vB$ is the (transverse) magnetic field and $\vD = \vD_{\perp}$ is the (transverse for a neutral system) displacement field. Here we ignored magnetic and diamagnetic interactions with the material system. $\vD$ and $\vP$ are related to the electric field according to $\vD = \epsilon_0 \vE_{\perp} + \vP_{\perp}$  and $\epsilon_0\vE_{\parallel} + \vP_{\parallel} = 0$.  We also note that the dipole-dipole interactions between any two molecules can be written in terms of the longitudinal polarizations associated with these molecules
	\begin{equation}
	V_{\text{Coul}}^{(nl)} = \frac{1}{\epsilon_0} \int d\vR \vP_{\parallel}^{(n)}(\vR)\cdot \vP_{\parallel}^{(l)}(\vR) \ \ \ (\text{for\ } n \neq l)
	\end{equation}
	so that, using $\vP(\vR) = \sum\limits_{n=1} \vP^{(n)}(\vR) = \sum\limits_{n=1}\left( \vP_{\perp}^{(n)}(\vR) + \vP_{\parallel}^{(n)}(\vR)\right)$, and assuming we operate in the point-dipole approximation  $\vP^{(n)}(\vR) = \vmu^{(n)}\delta(\vR - \vR_n)$, we thus find
	\begin{equation}\label{eq:cancellation}
	\begin{aligned}
	&\sum_{n < l} V_{\text{Coul}}^{(nl)} + \frac{1}{2\epsilon_0} \int d\vR \vP_{\perp}(\vR)\cdot \vP_{\perp}(\vR) \\
	= \ & \frac{1}{\epsilon_0} \sum_{n < l} \int d\vR \vP^{(n)}(\vR)\cdot \vP^{(l)}(\vR) \\
	& + 
	\frac{1}{2\epsilon_0} \sum_{n}\int d\vR \vP_{\perp}^{(n)}(\vR)\cdot \vP_{\perp}^{(n)}(\vR)
	\end{aligned}
	\end{equation}
	The first term on the right of Eq. \eqref{eq:cancellation} vanishes by the  assumption in our model that the charge distributions associated with different molecules do not overlap, which also reflects the retarded nature of light-matter interactions. Note that such a cancellation is valid  both in free space and in cavities\cite{Power1982}. The Hamiltonian \eqref{eq:H_PZW} then becomes
	\begin{equation}\label{eq:H_PZW2}
	\begin{aligned}
	H = \ & \sum_{n = 1}^{N} H_n +\frac{1}{2\epsilon_0} \sum_{n} \int d\vR \vP^{(n)}_{\perp}(\vR)\cdot \vP^{(n)}_{\perp}(\vR) \\
	&
	+ \frac{1}{2} \int d \vR \left(\frac{1}{\epsilon_0} \vD(\vR)\cdot \vD(\vR) + \frac{1}{\mu_0} \vB(\vR)\cdot \vB(\vR)\right) \\
	&
	- \frac{1}{\epsilon_0}\sum_{n = 1}^{N}\int d\vR \vD(\vR)\cdot \vP^{(n)}_{\perp}(\vR) 
	\end{aligned}
	\end{equation}
	At this point, we explicitly assume the field is classical. We write $\vD(\vR) = \sum \limits_{\mathbf{k}, \vxi} i\sqrt{\frac{\hbar \omega_k \epsilon_0}{2}}(a_k \mathbf{f}_k(\vR) - a_k^{\ast}\mathbf{f}^{\ast}_k(\vR))$, $a$ ($a^{\ast}$) are the classical analogs of the quantum annihilation (creation) operators and
	$\mathbf{f}_k(\vR)$ denotes the mode function which satisfies the Helmholtz equation with a certain boundary condition due to the cavity\cite{Power1982}. For example, in free space, $\mathbf{f}_k(\vR) = \vxi \frac{1}{\sqrt{\Omega}} e^{i \vk \cdot \vR}$.

	To arrive at a convenient Hamiltonian for studying the cavity effect, we rewrite the spatial integrations in Eq. \eqref{eq:H_PZW} by the corresponding reciprocal-space integrations. For example, the free-field part is $\frac{1}{2} \int d \vR \left(\frac{1}{\epsilon_0} \vD(\vR)\cdot \vD(\vR) + \frac{1}{\mu_0} \vB(\vR)\cdot \vB(\vR)\right) = \sum\limits_{\mathbf{k}, \vxi} \hbar\omega_k a^{\ast}_k a_k$. 
	For point dipoles such as $\vP^{(n)}(\vR) = \vmu^{(n)}\delta(\vR - \vR_n)$, we have
	$\int d\vR \vD(\vR)\cdot \vP^{(n)}_{\perp}(\vR) = \int d\vR \vD(\vR)\cdot \vP^{(n)}(\vR) = \sum \limits_{\mathbf{k}, \vxi} i\sqrt{\frac{\hbar \omega_k \epsilon_0}{2}}(a_k \mathbf{f}_k(\vR_n) - a_k^{\ast}\mathbf{f}^{\ast}_k(\vR_n))\cdot \vmu^{(n)}$. Likewise,
	$\int d\vR \vP^{(n)}_{\perp}(\vR)\cdot \vP^{(n)}_{\perp}(\vR) = \int d\vR d\vR'  \vP^{(n)}(\vR) \cdot \overset{\leftrightarrow}{\boldsymbol{\delta}}_{\perp}^{\red{\text{cav}}}(\vR - \vR') \vP^{(n)}(\vR')$; because the transverse $\delta$-function \red{inside the cavity} is a rank-two tensor defined as $\overset{\leftrightarrow}{\boldsymbol{\delta}}_{\perp}^{\red{\text{cav}}}(\vR - \vR') \equiv \sum\limits_{\mathbf{k}, \vxi} \mathbf{f}_k(\vR)\mathbf{f}^{\ast}_k(\vR')$ (where we take outer product of $\mathbf{f}_k(\vR)$ and $\mathbf{f}^{\ast}_k(\vR')$), 
	$\int d\vR \vP^{(n)}_{\perp}(\vR)\cdot \vP^{(n)}_{\perp}(\vR) = \sum\limits_{\mathbf{k}, \vxi} \left | \int d\vR\vP^{(n)}(\vR) \mathbf{f}_k(\vR)\right |^2 = \sum\limits_{\mathbf{k}, \vxi} |\vmu^{(n)}\cdot \mathbf{f}_k(\vR_n)|^2$.
	Therefore, we can rewrite Eq. \eqref{eq:H_PZW2} as
	\begin{widetext}
	\begin{equation}\label{eq:PWZ_cavity}
	\begin{aligned}
		H &= \sum_{n = 1}^{N} H_n + \sum_{\mathbf{k} \in \text{supp}, \vxi} \left[\hbar\omega_k a^{\ast}_k a_k - \sum_{n = 1}^{N}  i\sqrt{\frac{\hbar \omega_k }{2\epsilon_0}}(a_k \mathbf{f}_k(\vR_n) - a_k^{\ast}\mathbf{f}^{\ast}_k(\vR_n))\cdot \vmu^{(n)} +   \sum_{n = 1}^{N}  \frac{1}{2\epsilon_0} |\vmu^{(n)}\cdot \mathbf{f}_k(\vR_n)|^2\right] \\
	\end{aligned}
	\end{equation}
	\end{widetext}
	where $\mathbf{k} \in \text{supp}$ denotes that the summation includes all supported cavity modes. \red{Eq. \eqref{eq:PWZ_cavity} preserves causality exactly.}
	Equivalently, when the photon modes are expressed as a function of positions and momenta such as  $a_k = (2\hbar\omega_k)^{-1/2}(\omega_k q_k + ip_k)$, Eq. \eqref{eq:PWZ_cavity} reads
	\begin{widetext}
	\begin{equation}\label{eq:PWZ_cavity2}
		\begin{aligned}
		H &= \sum_{n = 1}^{N} H_n  
		- \sum_{\mathbf{k} \in \text{supp}, \vxi} \sum_{n \neq l}\frac{1}{2\epsilon_0} \left(\vmu^{(n)}\cdot \mathbf{f}_k(\vR_n)\right)\left(\vmu^{(l)}\cdot \mathbf{f}^{\ast}_k(\vR_l)\right) 
		\\
		& + \sum_{\mathbf{k} \in \text{supp}, \vxi} \left[
		\frac{1}{2}\omega_k^2 \left(q_k + \sum_{n = 1}^{N} \frac{1}{\sqrt{\epsilon_0}\omega_k}\vmu^{(n)}\cdot \text{Im}\left[ \mathbf{f}_k(\vR_n)\right] \right)^2  + \frac{1}{2}\left(p_k - \sum_{n = 1}^{N} \frac{1}{\sqrt{\epsilon_0}}\vmu^{(n)}\cdot \text{Re}\left[ \mathbf{f}_k(\vR_n)\right]\right)^2
		\right] 
		\end{aligned}
	\end{equation}
	\end{widetext}
	 Note that
	the second term in Eq. \eqref{eq:PWZ_cavity2}  is simply the modified dipole-dipole interaction between molecules in the cavity:
	\begin{equation}\label{eq:V_dd_cavity}
	V_{\text{dd}}^{(nl)}(\Omega) = - \sum_{\mathbf{k} \in \text{supp}, \vxi} \frac{1}{2\epsilon_0} \left(\vmu^{(n)}\cdot \mathbf{f}_k(\vR_n)\right)\left(\vmu^{(l)}\cdot \mathbf{f}^{\ast}_k(\vR_l)\right) + \text{c.c.}
	\end{equation}
	which is a function of the cavity volume $\Omega$, and where c.c. denotes the complex conjugate. When $\Omega \rightarrow +\infty$, Eq. \eqref{eq:V_dd_cavity} reduces to the familiar free-space form, \red{$V_{\text{dd}}^{(nl)} = -\frac{1}{\epsilon_0} \vmu_{n}\cdot \overset{\leftrightarrow}{\boldsymbol{\delta}}_{\perp}(\vR_n - \vR_l) \vmu_{l} = \frac{1}{\epsilon_0} \vmu_{n}\cdot \overset{\leftrightarrow}{\boldsymbol{\delta}}_{\parallel}(\vR_n - \vR_l) \vmu_{l}$, where $\overset{\leftrightarrow}{\boldsymbol{\delta}}_{\perp}$ and $\overset{\leftrightarrow}{\boldsymbol{\delta}}_{\parallel}$ denotes the transverse and the longitudinal $\delta$-function in free space}.
	
	When Eq. \eqref{eq:PWZ_cavity2} is used for calculating the PMF \red{according to Eq. \eqref{eq:PMF}} (as what we did above),  we see again, that after integration over the radiation field \red{DoFs} ($p_k$ and $q_k$), the only possible source of many-molecular contributions to the single molecule PMF is the remaining dipolar interaction terms represented by $\sum\limits_{n < l} V_{\text{dd}}^{(nl)}(\Omega)$.
	\red{In other words, the only way the cavity can exert a meaningful effect on a molecule is by dressing or modifying the dipole-dipole interactions, presumably through image charges, etc. We will discuss the size of these modifications in the next section.  
		
	Before concluding this section, however, we note that all of the arguments so far have assumed a classical treatment of the nuclei and photons. In the Appendix we will show that,  if the quantum nature of the nuclei and photons are considered, a path-integral calculation does lead to an additional modification of the PMF (see Eqs. \eqref{eq:PMF_rpmd_exact} and \eqref{eq:V_eff}). However, we find that it is difficult to interpret the nature of this term even qualitatively with complete rigor, and future research will likely need to invoke numerical simulations with realistic parameters.}

		\begin{figure}
	\includegraphics[width=1.0\linewidth]{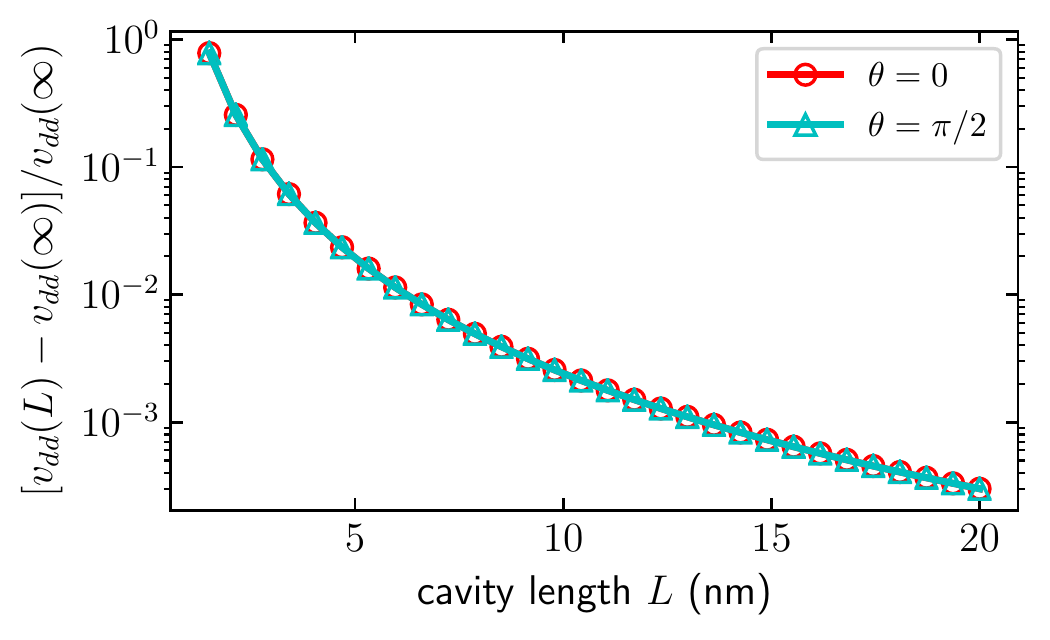}
	\caption{
		The normalized difference between the dipole-dipole interaction inside a cavity and that in free space, Eq. \eqref{eq:eta}, plotted against the cavity length  $L$.
		The cavity mirrors are set at $z = 0$ and $z = L$ and two paralleled dipoles are located at $(0, 0, \frac{L + \Delta r}{2})$ and $(0, 0, \frac{L - \Delta r}{2})$ with an angle $\theta$ with respect to the $z$ axis. The separation between the dipoles is set as $\Delta r = 1$ nm.
		Note that the dimensionless quantity $[v_{\text{dd}}(L) - v_{\text{dd}}(\infty)] / v_{\text{dd}}(\infty)$ shows no dependence on the angle $\theta$.}
	\label{fig:ratio}
\end{figure}
	
	\section{Discussion}
	\label{sec:discussion}
	The analysis carried in the previous \red{section} indicates that, for a system of $N$ molecules that are coupled to each other through their interaction with  cavity modes, the only possible source of "collective" ($N$ dependent) effect at the single-molecular \red{classical} TST rate stems from the cavity effect on the intermolecular (assumed to be dominated by dipolar) interactions. Such effects were investigated in the past, where in the electrostatic (long-wave) limit they can be described in terms of interaction of a given molecular dipole with the infinite number of images associated with each molecule positioned between cavity mirrors\cite{Takae2013,DeBernardis2018,Kobayashi1995,Hopmeier1999,Goldstein1997}. To estimate the magnitude of this effect we consider a cavity bounded by perfect mirrors located at $z=0$   and  $z=L$, in which two point dipoles are positioned at  $(0, 0, \frac{L + \Delta r}{2})$ and  $(0, 0, \frac{L - \Delta r}{2})$. An analytical expression for the dipole-dipole interaction in such configuration is provided in Ref. \cite{Takae2013} (see Eq. (3.9) therein). Fig. \ref{fig:ratio} plots the normalized difference between the dipole-dipole interaction ($v_{\text{dd}}$)  inside the cavity and in free space,
	\begin{equation}\label{eq:eta}
	\eta(L) \equiv \frac{v_{\text{dd}}(L) - v_{\text{dd}}(\infty)}{v_{\text{dd}}(\infty)}
	\end{equation}
	as a function of the cavity length  $L$. It is seen that when the cavity length is comparable with the separation between the dipoles (with $\Delta r = 1$  nm), the dipole-dipole interaction is affected significantly. 
	\red{Therefore, one may expect that for molecules located close to the cavity mirrors, the PMF of a single molecule can be modified because of the modification of intermolecular electrostatic interactions by the mirror as well as the trend for the molecules to collectively orient relative to the mirror, as recently demonstrated in Ref. \citenum{Galego2019} (using cavity lengths $< 10$ nm).} However, for cavities usually used in
	studies of VSC \red{catalysis} (with the cavity length of microns), \red{the fraction of molecules in proximity
	to the cavity mirrors is not meaningful, nor are the molecules anisotropically oriented in liquid-phase reactions. Hence, one can conclude that} the cavity effect on intermolecular dipole interaction is
	negligible \red{for typical liquid-phase VSC experiments}.

	We \red{conclude that, the} cavity effects on  the rate of chemical processes whose underlying nuclear dynamics takes place on the ground electronic potential surface  cannot be accounted for \red{at the level of classical transition state theory}  where the rate is determined by the thermal distribution of nuclear configurations on the ground-state potential surface.
	While proximity to mirrors can affect intermolecular interactions, these effects are negligible for standard cavities used in VSC studies with $L$ of the order of microns. \red{This state of affairs then leaves us two options.  First, quantum-mechanical features of the nuclei and photons could perhaps lead to important modifications of the PMF (though our hypothesis for now is that this is unlikely; see below). Second,  experimentally} observed cavity effects on ground (electronic) state reactions \red{can also} arise from other physical effects, that may have classical (excitation rates or barrier crossing efficiencies) or quantum (e.g. non-adiabatic transitions) origins.

		\begin{appendices}
		
	\red{
		
	\section{Path Integral Treatment}

		\setcounter{equation}{0}
		\setcounter{figure}{0}
		\setcounter{table}{0}
		\renewcommand{\theequation}{A\arabic{equation}}
		\renewcommand{\thefigure}{A\arabic{figure}}

	The calculation of the PMF above relies on an important assumption: the classical treatment of nuclei and photons. At this point, a natural question arises: will the quantum effects of nuclei and photons contribute to the PMF change? 
	In fact, quantum effects may be relevant for VSC experiments where the  resonant mode frequency ($800 \sim 4000 \text{\ cm}^{-1}$) is much larger than room temperature ($\sim 200 \text{\ cm}^{-1}$). Therefore, it is necessary to investigate the PMF change by considering  quantum nuclei and photons.
	
	In molecular dynamics simulations, the standard way to consider the quantum effects for nuclei is to perform path-integral calculations. For the system we are interested in (nuclei + photons in the electronic ground state), below we will perform a path-integral calculation for the coupled photon-nuclear system characterized by the quantum Hamiltonian \eqref{eq:H'2}. Because the photon mode is a harmonic oscillator, we can rewrite Eq. \eqref{eq:H'2} as
	\begin{equation}\label{eq:H'3}
	\begin{aligned}
	\hat{H} &=  \sum_{i} \frac{\hat{\mathbf{p}}_i^2}{2 m_i} +  \hat{V}_{\text{Coul}}\left(\{\hat{\vR}_i\}\right) \\
	&
	+ \frac{1}{2}\omega^2 \left(\hat{q} + \frac{1}{\sqrt{\Omega\epsilon_0} \omega}\left(\hat{\vmu}_S\cdot \vxi \right) \right)^2 + \frac{1}{2}\hat{p}^2
	\end{aligned}
	\end{equation}
	where the potential is a function of coordinates only. Follow the derivations in Eqs. \eqref{eq:PES} and \eqref{eq:Treatment_2}, after projecting the above Hamiltonian to the electronic ground state according to the Hartree approximation, Eq. \eqref{eq:H'3} becomes
	\begin{subequations}\label{eq:H'4}
	\begin{align}
	\hat{H}^{\text{G}} &=  
	\hat{H}^{\text{G}}_{\text{0}} + \hat{H}^{\text{G}}_{\text{F}}
	\end{align}
	Here, the nuclear part is defined as
	\begin{align}
	\hat{H}^{\text{G}}_{\text{0}} = \sum_{i=1}^{N_{\text{nuc}}} \frac{\hat{\mathbf{P}}_i^2}{2 M_i} +  \hat{V}_{\text{Coul}}\left(\{\hat{\mathbf{R}}_i\}\right)
	\end{align}
	The field-related part is
	\begin{align}
	\hat{H}^{\text{G}}_{\text{F}} = \frac{1}{2}\hat{p}^2 + \frac{1}{2}\omega^2 \left(\hat{q} + \frac{1}{\sqrt{\Omega\epsilon_0} \omega} \sum_{n=1}^{N} \hat{d}_{ng}\right)^2
	\end{align}
	where $\hat{d}_{ng} \equiv \sum_{n} \braket3{\psi_{ng}}{\hat{\vmu}_n}{\psi_{ng}} \cdot \vxi$ denotes the dipole moment operator for the $n$-th molecule in the electronic ground state.
	\end{subequations}
	Note that the Hamiltonian \eqref{eq:H'4} contains the operators for both nuclei ($\hat{\mathbf{P}}_i$, $\hat{\mathbf{R}}_i$) and photons ($\hat{p}$, $\hat{q}$) and we have neglected the self-dipole fluctuations term (the last term in Eq. \eqref{eq:Treatment_2}).

	 The quantum canonical partition function ($\mathcal{Z}$) for Hamiltonian \eqref{eq:H'4} reads:
	\begin{equation}
	\mathcal{Z} = \trtwo{np}{e^{-\beta \hH^{\text{G}}}} 
	\end{equation}
	where $\trtwo{np}{\cdots}$ denotes the trace over both the nuclear and photonic DoFs. For performing a path-integral expansion \cite{Tuckerman2010,Chowdhury2017},  we use a Trotter expansion such as
	$ e^{-\beta \hH} = \lim\limits_{M\rightarrow +\infty} \prod_{\alpha=1}^{M}e^{-\beta_M \hat{V}/2} \allowbreak  e^{-\beta_M \hat{T}} e^{-\beta_M \hat{V}/2}$. Here, we have split the exponential into $M$ slices and $\beta_{M} \equiv \beta / M$. $\hat{T}$ and $\hat{V}$ denote the kinetic and the potential parts of the coupled photon-nuclear system. After inserting $M$ copies of the resolution of identity $I_{\mathbf{R}} = \int d \mathbf{R}_i^{(\alpha)} \ket{\mathbf{R}_i^{(\alpha)}} \bra{\mathbf{R}_i^{(\alpha)}} \int d q^{(\alpha)} \ket{q^{(\alpha)}}\bra{q^{(\alpha)}}$ and $I_{\mathbf{P}} = \int d \mathbf{P}_i^{(\alpha)} \ket{\mathbf{P}_i^{(\alpha)}} \bra{\mathbf{P}_i^{(\alpha)}} \int d p^{(\alpha)} \ket{p^{(\alpha)}}\bra{p^{(\alpha)}}$, we obtain a classical isomorphism for the partition function:
	\begin{equation}
	\begin{aligned}
	\mathcal{Z} &=   \lim_{M \rightarrow +\infty}  \frac{1}{(2\pi \hbar)^f} \\
	&\times \int d \{\mathbf{P}_i^{(\alpha)}\} d \{\mathbf{R}_i^{(\alpha)}\} d \{p^{(\alpha)}\} d \{q^{(\alpha)}\} e^{-\beta_M H^{\text{rp}}} 
	\end{aligned}
	\end{equation}
	where $f = M (N_{\text{nuc}} + 1)$ and $H^{\text{rp}}$ is the classical ring-polymer Hamiltonian for the coupled photon-nuclear system:
	\begin{equation}
	H^{\text{rp}} = H_{0}^{\text{rp}} +  H_{F}^{\text{rp}}
	\end{equation}
	Here, $H_0^{\text{rp}}$ denotes the conventional ring-polymer Hamiltonian for the nuclear part:
	\begin{equation}\label{eq:H0_RPMD}
	\begin{aligned}
	H_0^{\text{rp}} &= \sum_{\alpha=1}^{M} \sum_{i=1}^{N_{\text{nuc}}} \Bigg \{ \frac{\left[\mathbf{P}_i^{(\alpha)}\right]^2}{2 M_i} + V_{\text{Coul}}(\{\mathbf{R}_i^{(\alpha)}\}) \\
	&+ \frac{1}{2}M_i \omega_M^2 \left[\mathbf{R}_i^{(\alpha)} - \mathbf{R}_i^{(\alpha-1)}\right]^2 \Bigg \}
	\end{aligned}
	\end{equation}
	where $\omega_M \equiv 1 / \beta_M \hbar$ and $\mathbf{R}_i^{(0)} = \mathbf{R}_i^{(M)}$. In Eq. \eqref{eq:H0_RPMD}, $M$ copies (aka $M$ beads) of the classical nuclear Hamiltonian (the first two terms above) are coupled together by spring constants that capture  the quantum effects of nuclei. Similarly,
	$H_{F}^{\text{rp}}$  denotes the ring-polymer Hamiltonian for the photon-related part:
	\begin{equation}
	\begin{aligned}
			H_{F}^{\text{rp}} &= \sum_{\alpha=1}^{M} 
			\Bigg \{
			\frac{1}{2}\left[p^{(\alpha)}\right]^2 + \frac{1}{2}\omega^2 \left [ q^{(\alpha)} + \frac{1}{\sqrt{\Omega \epsilon_0} \omega}\sum_{n=1}^{N}d_{ng}^{(\alpha)}\right]^2 \\
			& + \frac{1}{2}\omega_M^2 \left [q^{(\alpha)} - q^{(\alpha-1)}\right]^2 \Bigg \}
	\end{aligned}
	\end{equation}
	where $q^{(0)} = q^{(M)}$.
	Note that, because the ground-state dipole moment for molecule $n$ ($d_{ng}^{(\alpha)}$) is a function of nuclear coordinates, $d_{ng}^{(\alpha)}$ can take different values for different beads.
	
	If $\hat{A}$ is a function of spatial coordinates only, the thermal average 
	\begin{equation}
	\avg{\hat{A}} = \tr{\hat{A}e^{-\beta \hH^{\text{G}}}}
	\end{equation}
	can be calculated by the following classical phase-space average:
	\begin{equation}
	\begin{aligned}
	&\avg{\hat{A}(\{\hat{\mathbf{R}}_i, \hat{q}\})} = \lim_{M \rightarrow +\infty}  \frac{1}{(2\pi \hbar)^f \mathcal{Z}} \\
	&\times \int d \{\mathbf{P}_i^{(\alpha)}\} d \{\mathbf{R}_i^{(\alpha)}\}  d \{p^{(\alpha)}\} d \{q^{(\alpha)}\} 
	e^{-\beta_M H^{\text{rp}}} \mathcal{A}_M
	\end{aligned}
	\end{equation}
	where
	\begin{equation}
	\mathcal{A}_M = \frac{1}{M} \sum_{\alpha=1}^{M} A(\{\mathbf{R}_i^{(\alpha)}, q^{(\alpha)}\})
	\end{equation}
	In order to calculate the PMF along a putative reaction pathway, let us make the simple assumption that $\hat{\mathbf{R}}_j$ is a reasonable reaction coordinate. Then,
	\begin{equation}
	\mathcal{A}_M = \frac{1}{M} \sum_{\alpha'=1}^{M} \delta(\mathbf{R}_j - \mathbf{R}_j^{(\alpha')})
	\end{equation}
	and the PMF is calculated by
		\begin{equation}
		\begin{aligned}
		e^{-\beta F(\mathbf{R}_j)} &= \lim_{M \rightarrow +\infty}  \frac{1}{M} \sum_{\alpha'=1}^{M} \frac{1}{(2\pi \hbar)^f \mathcal{Z}}  \\
		&\times \int d \{\mathbf{P}_i^{(\alpha)}\} d '\{\mathbf{R}_i^{(\alpha)}\} d \{p^{(\alpha)}\} d \{q^{(\alpha)}\} e^{-\beta_M H^{\text{rp}}} 
		\end{aligned}
	\end{equation}
	where $d '\{\mathbf{R}_i^{(\alpha)}\}$ denotes excluding the integral over $\mathbf{R}_j^{(\alpha')}$.
	
	To proceed with our calculation, let us make the replacement  $q^{(\alpha)}  \leftarrow q^{(\alpha)} + \frac{1}{\sqrt{\Omega \epsilon_0} \omega}\sum_{n=1}^{N}d_{ng}^{(\alpha)}$, and so we  find the above integral is equivalent to
		\begin{equation}\label{eq:PMF_rpmd}
		\begin{aligned}
		&e^{-\beta F(\mathbf{R}_j)} = \lim_{M \rightarrow +\infty}  \frac{1}{M} \sum_{\alpha'=1}^{M} \frac{1}{(2\pi \hbar)^f \mathcal{Z}} \\
		& \times \int d \{\mathbf{P}_i^{(\alpha)}\} d '\{\mathbf{R}_i^{(\alpha)}\} d \{p^{(\alpha)}\} d \{q^{(\alpha)}\}  e^{-\beta_M \left(H^{\text{rp}}_0 + H^{\text{rp}}_{F0} + V^{\text{rp}} \right)} 
		\end{aligned}
	\end{equation}
	where $H^{\text{rp}}_0$ is defined in Eq. \eqref{eq:H0_RPMD}, $H^{\text{rp}}_{F0}$ denotes the ring-polymer Hamiltonian for a free photon:
	\begin{equation}
	\begin{aligned}
	H_{F0}^{\text{rp}} &= \sum_{\alpha=1}^{M} \frac{1}{2}\left[ p^{(\alpha)} \right]^2 + \frac{1}{2}\omega^2 \left[q^{(\alpha)}\right]^2 \\
	&+ \frac{1}{2}\omega_M^2 \left [q^{(\alpha)} - q^{(\alpha-1)}\right]^2
	\end{aligned}
	\end{equation}
	and $V^{\text{rp}}$ arises from the quantum interbead interactions between nuclei and photons
	\begin{equation}
	\begin{aligned}
	V^{\text{rp}} &= \sum_{\alpha=1}^{M} -\frac{\omega_M^2}{\sqrt{\Omega \epsilon_0} \omega} \left [q^{(\alpha)} - q^{(\alpha-1)}\right] \left [ \sum_{n=1}^{N} d_{ng}^{(\alpha)} - d_{ng}^{(\alpha-1)} \right]  \\
	&+ \frac{\omega_M^2}{2\Omega \epsilon_0 \omega^2} \left [ \sum_{n=1}^{N} d_{ng}^{(\alpha)} - d_{ng}^{(\alpha-1)} \right]^2
	\end{aligned}
	\end{equation}
	
	\subsection{Exact Solution}
	To further identify the effect of $V^{\text{rp}}$ on the PMF, let us formally integrate out the photonic DoFs using the following identity:
	\begin{equation}
	\int d\{q^{(\alpha)}\} \ e^{-\frac{1}{2} \vec{q}^{T} \mathbf{A} \vec{q} + \vec{b}^T \vec{q}} = 
	\sqrt{\frac{(2\pi)^M}{\det\mathbf{A}}} e^{\frac{1}{2}\vec{b}^{T}\mathbf{A}^{-1}\vec{b}}
	\end{equation}
	where $\vec{q} = [q^{(1)}, q^{(2)}, \cdots, q^{(M)}]$. Given the definition of matrix $\mathbf{A}$ and vector $\vec{b}$ as
	\begin{equation}
	\mathbf{A}_{\alpha,\alpha'}
	=
	\begin{dcases}
	\beta_{M}\left( \omega^2 + 2\omega_M^2 \right),& \text{if } \alpha = \alpha'\\
	-2 \beta_{M} \omega_M^2,              & \text{if } \alpha = \alpha'\pm 1\\
	0,              & \text{otherwise} 
	\end{dcases}
	\end{equation}
	and
	\begin{equation}
	\vec{b}_{\alpha} = \beta_{M} \frac{\omega_M^2}{\sqrt{\Omega \epsilon_0} \omega} \left [ \sum_{n=1}^{N} 2d_{ng}^{(\alpha)} - d_{ng}^{(\alpha-1)} - d_{ng}^{(\alpha+1)} \right] 
	\end{equation}
	after the integration over the photonic DoFs, the PMF in Eq. \eqref{eq:PMF_rpmd} is equivalent to
	\begin{equation}\label{eq:PMF_rpmd_exact}
	\begin{aligned}
	e^{-\beta F(\mathbf{R}_j)} &= \lim_{M \rightarrow +\infty}  \frac{1}{M} \sum_{\alpha'=1}^{M} \frac{1}{(2\pi \hbar)^f \mathcal{Z}_0} \\
	&\times \int d \{\mathbf{P}_i^{(\alpha)}\} d '\{\mathbf{R}_i^{(\alpha)}\} e^{-\beta_M \left(H^{\text{rp}}_0 + V_{\text{eff}} \right)} 
	\end{aligned}
	\end{equation}
	Here, $\mathcal{Z}_0 =   \lim\limits_{M \rightarrow +\infty}  (2\pi \hbar)^{-MN_{\text{nuc}}} \int d \{\mathbf{P}_i^{(\alpha)}\} d \{\mathbf{R}_i^{(\alpha)}\}  e^{-\beta_M H_0^{\text{rp}}}$ denotes the partition function without the cavity mode.  
	$V_{\text{eff}}$ represents the effective cavity modification due to the quantum effects of nuclei and photons and is a function of the nuclear DoFs only:
	\begin{equation}\label{eq:V_eff}
	\begin{aligned}
	V_{\text{eff}} &= \sum_{\alpha=1}^{M} \frac{\omega_M^2}{2\Omega \epsilon_0 \omega^2} \left [ \sum_{n=1}^{N} d_{ng}^{(\alpha)} - d_{ng}^{(\alpha-1)} \right]^2 + \frac{1}{2\beta_{M}} \vec{b}^{T}\mathbf{A}^{-1}\vec{b} \\
	&- \frac{1}{\beta_{M}} \ln\left(\sqrt{\frac{(\beta_{M}\omega^2)^M}{\det\mathbf{A}}}\right)
	\end{aligned}
	\end{equation}
	
	In general, it is difficult to interpret the PMF constituted by Eqs. \eqref{eq:PMF_rpmd_exact} and \eqref{eq:V_eff}
	\red{as we are not aware of any symmetry we can use to state definitively whether or not (and if so how) $V_{\text{eff}}$ will depend on $N$ (the number of molecules); future numerical work will be necessary to address this point. That being said, before} ending this paper, we note that for VSC experiments, the catalytic effect usually vanishes when the detuning (i.e., the frequency difference between the cavity mode and the vibrational frequency) increases \cite{Thomas2016,Lather2019,Hiura2018,Thomas2019_science}. In Eq. \eqref{eq:V_eff}, however, $V_{\text{eff}}$ does not seem to show such a delta-function-like dependence on the detuning ($\omega - \omega_{0}$).
	Therefore, at this point, even without any numerical evidence, one may hypothesize that the quantum modification $V_{\text{eff}}$ on the PMF may not be responsible for VSC catalysis, but this hypothesis needs further numerical verification.

}

	\end{appendices}

	\section{Acknowledgements}\label{sec:acknowledgement}
	This material is based upon work supported by the U.S. Department of Energy, Office of Science, Office of Basic Energy Sciences under Award Number DE-SC0019397 (J.E.S.),
	U.S. National Science Foundation, Grant No. CHE1665291(A.N.) and the Israel-U.S. Binational Science Foundation, Grant No. 2014113 (A.N.). T.E.L. also acknowledges the Vagelos Institute for Energy Science and Technology at the University of Pennsylvania for a graduate fellowship. 
	This research also used resources of the National Energy Research Scientific Computing Center (NERSC), a U.S. Department of Energy Office of Science User Facility operated under Contract No. DE-AC02-05CH11231.  
	We  thank A. Semenov and M. Dinpajooh for helpful discussions.
	
	\section{Data Availability Statement}                                                   Data available in article or supplementary material.     
	

\begin{thebibliography}{42}
\bibitem [{foo()}]{footnote1}%
		\BibitemOpen
		\href@noop {} {}\bibinfo {howpublished} {Note that even if the molecule does
			not have a permanent dipole moment in its equilibrium configuration, a dipole
			may appear due to distortion resulting from nuclear motion.}\BibitemShut
		{Stop}%
\end{thebibliography}
	
	%

\end{document}